# pCoRev: Physics-Informed Contrastive Learning for Unsupervised Reverberation Artifact Suppression in Ultrafast Plane-Wave Imaging

Hyunwoo Cho, Yeeun Kwon, Yangmo Yoo, *and* Jinbum Kang

***Abstract*—Reverberation artifact suppression in ultrafast plane-wave (PW) imaging remains challenging because artifact-free reference data are generally unavailable and multipath echoes often resemble tissue signals. Coherent plane-wave compounding (CPWC) attenuates reverberation only partially and sacrifices the frame rate that motivates PW imaging. We propose pCoRev, an unsupervised physics-informed contrastive learning framework that leverages the distinct angular coherence characteristics of tissue echoes and reverberation artifacts. A single-angle I/Q frame is the network input, and the remaining angles of the same CPWC acquisition provide angle-consistent positive pairs, residual-derived negative pairs, and coherence-guided self-supervision, so no paired artifact-free data are required. On a fabricated reverberation phantom with 31 validation acquisitions, pCoRev achieved the best contrast ratio (CR), contrast-to-noise ratio (CNR), and generalized CNR (gCNR) of 38.76 ± 5.40 dB, 11.75 ± 1.81 dB, and 0.932 ± 0.051 against delay-and-sum (DAS), delay-multiply-and-sum, a supervised 3D convolutional neural network, and unsupervised deep coherence learning. Because the network is trained on single-angle inputs, it also operates on a single PW transmission: with one angle, pCoRev reached a CR of 32.10 ± 7.79 dB and a CNR of 10.39 ± 3.02 dB, exceeding 75-angle compounded DAS (20.45 ± 3.84 dB and 8.38 ± 2.00 dB) with 75-fold fewer transmissions. On four clinical acquisitions from other scanners, pCoRev improved CR over DAS by 9–21 dB without fine-tuning. These results indicate that physics-informed unsupervised contrastive learning suppresses reverberation effectively while preserving the frame-rate advantage of ultrafast PW imaging.**

## I. Introduction

Medical ultrasound (US) imaging is often described under the working assumption that each detected echo arises from a single, direct transit of an acoustic pulse from the transducer to a scatterer and back [1]. In clinical settings this assumption is frequently violated [2]: at interfaces with a large impedance mismatch, such as bone, air, or calcification, the acoustic energy undergoes multiple reflections before returning to the probe, which produces reverberation artifacts [3-6]. By overlaying plausible but false information onto the image, reverberation can obscure subtle lesions and, in some cases, mimic pathology, thereby complicating interpretation [6-8]. Clinical examples include valve-related reverberations mistaken for thrombi, capsule echoes obscuring renal calculi, and clutter that alters the apparent extent of atherosclerotic plaques [9-11].

Suppressing reverberation is nontrivial for three reasons. Reverberations stem from the same transmit events as the desired echoes and share their spectral bandwidth and phase characteristics [6, 8, 12], which limits frequency-domain filtering. Their spatial distribution and intensity are nonstationary, varying with small changes in probe pressure, respiration, or steering angle, so methods that assume stationary statistics or fixed parameters are unreliable [8]. Multi-path echoes also remain temporally coherent with tissue motion, which defeats the frame averaging and temporal filtering that work against uncorrelated noise [13, 14]. A practical strategy must therefore discriminate single- and multi-path echoes from spatial and contextual cues within a single acquisition [15].

These difficulties are amplified in ultrafast plane-wave (PW) imaging. Unfocused PW transmissions enable frame rates of several kilohertz and underpin shear-wave elastography, ultrafast Doppler, and functional ultrasound [16, 17], but a single transmission insonifies the entire field of view, so multipath echoes from superficial layers contaminate the whole image. Coherent plane-wave compounding (CPWC) recovers image quality by averaging tens of steered transmissions [16], which partly decorrelates angle-dependent reverberation, but it divides the frame rate by the number of angles. Suppressing reverberation from one or a few transmissions is therefore a prerequisite for retaining the ultrafast frame rate.

Manuscript received October 26, 2026;

This research was supported by the “Regional Innovation System & Education (RISE)” program through the Seoul RISE Center, funded by the Ministry of Education (MOE) and the Seoul Metropolitan Government (2026-RISE-01-013-05). This work was also supported by the National Research Foundation of Korea (NRF) grant funded by the Korea government (MSIT) (RS-2025-00513383). This research was also supported by Kyungpook National University Research Fund, 2026.

Hyunwoo Cho is with the School of Electronics Engineering, Kyungpook National University, Daegu 41566, Korea (e-mail: hwcho@knu.ac.kr).

Yeeun Kwon is with the Department of Electronic Engineering, Sogang University, Seoul 04107, Korea (e-mail: ky1511@sogang.ac.kr)

Yangmo Yoo is with the Department of Electronic Engineering, Sogang University, Seoul 04107, Korea and with the Department of Biomedical Engineering, Sogang University, Seoul 04107, Korea. (e-mail: ymyoo@sogang.ac.kr)

Jinbum Kang is with the Department of Medical Artificial Intelligence Engineering, Dongguk University, Seoul 04620, Korea. (e-mail: jinbum.kang@dongguk.edu) (Corresponding author)

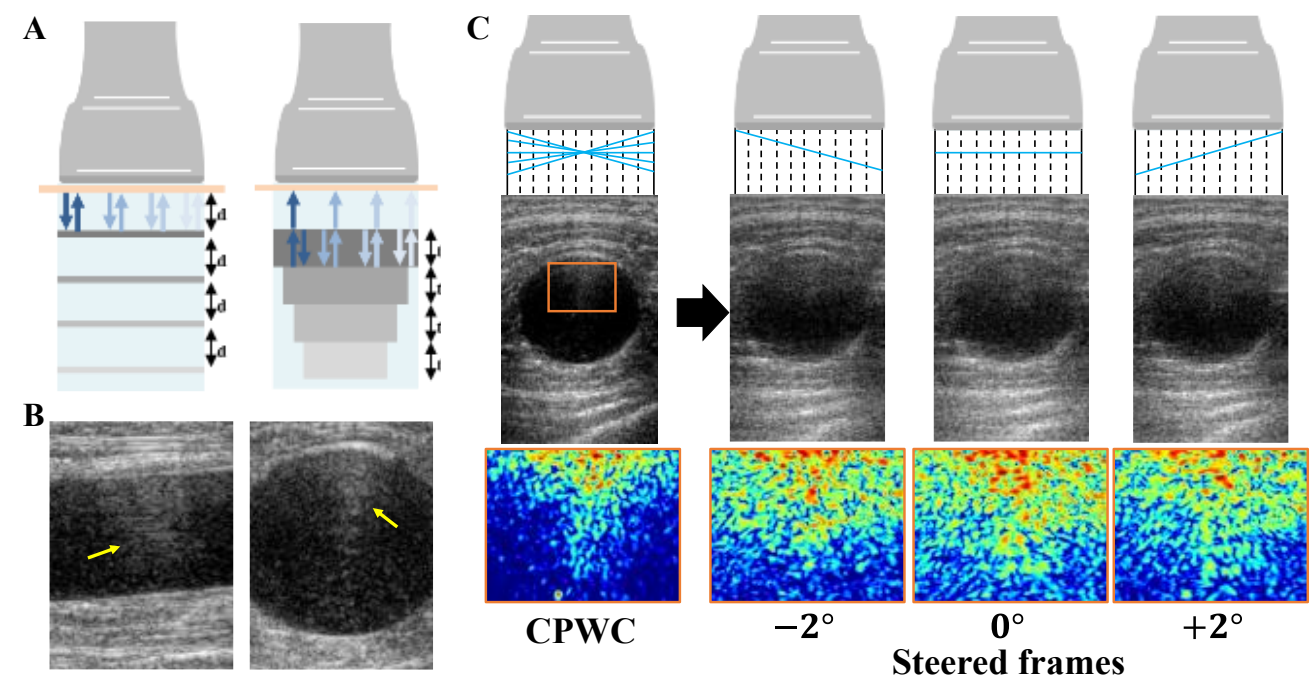

**Fig. 1** Physical principle and angular dependence of reverberation artifacts. (A) Reverberations caused by single-path or multi-path reflections. (B) Examples of reverberation artifacts in B-mode ultrasound. (C) In CPWC imaging, anatomy remains angle-consistent, whereas reverberation varies with steering angle, forming the basis of the proposed unsupervised framework.

Advanced beamformers, including coherence factor (CF) and short-lag spatial coherence (SLSC) based adaptive methods and delay-multiply-and-sum (DMAS), can reduce incoherent clutter or enhance contrast [18-20], but their performance may degrade when reverberation exhibits coherence similar to tissue echoes, and aggressive suppression may attenuate weak anatomical structures [13, 21].

Deep learning has been explored for ultrasound reconstruction and artifact suppression [18], including beamforming [19-21], speckle reduction [22, 23], and noise suppression [24, 25]. For reverberation, however, suitable training data are hard to obtain. Simulation-based studies mostly address diffuse clutter, because widely used simulators such as Field II [26] adopt single-scattering assumptions that do not capture multiple reflections, and the resulting sim-to-real gap limits generalization to clinical acquisitions [27]. Real-world training data avoid this gap, but paired artifact-present and artifact-free acquisitions are generally unobtainable for reverberation [28, 29].

Self-supervised and unsupervised reconstruction methods alleviate the need for labeled pairs: sparsity-promoting formulations for B-mode reconstruction [30], unsupervised speckle reduction exploiting the variability across steered PW transmissions or dedicated speckle simulation [28, 31, 32], and deep coherence learning (DCL), which reconstructs B-mode images by maximizing inter-angle coherence without labels [28]. These methods exploit physical structure rather than ground-truth labels, but none specifically separates multi-path reverberation from direct echoes. Contrastive learning (e.g., SimCLR [33], MoCo [34]) offers strong representation learning, yet its use in ultrasound has focused on image analysis rather than artifact suppression [35].

We propose pCoRev, a physics-informed contrastive learning framework for unsupervised reverberation suppression, motivated by the observation that anatomical echoes remain coherent across steering angles whereas multi-path reverberation varies with the angle. Using multi-angle I/Q acquisitions, the network is trained to move its prediction toward angle-consistent references and away from residual reverberation-like patterns, without labels or artifact-free targets. Because the network takes a single steered frame as input and uses the other angles only as self-supervisory keys, the trained model applies to a single PW transmission as well as to compounded data. We evaluate pCoRev against conventional and deep-learning methods on phantom and *in vivo* data and quantify its performance as the number of transmit angles is reduced from 75 to one.

## II. Materials and Methods

### A. Physical properties of reverberation artifacts

Medical ultrasound is commonly reconstructed under a single-path assumption. However, strong impedance boundaries can produce delayed multipath returns that are reconstructed at incorrect depths, generating reverberation artifacts such as repetitive lines, banding, or diffuse haze. As illustrated in Figs. 1A and 1B, these artifacts can overlap with true anatomy and reduce target conspicuity.

The key physical cue of this study is the different angular behavior of tissue echoes and multipath reverberation (Fig. 1C): in CPWC imaging, anatomical echoes remain spatially aligned across small steering-angle changes, whereas multipath reverberation shifts and decorrelates because its path depends on the steering geometry. This dichotomy enables self-supervised learning within a single CPWC acquisition: pCoRev is encouraged to preserve coherent tissue-like content while suppressing angle-dependent components of reverberation artifacts, without artifact-free targets.

### B. Contrastive learning for reverberation artifact suppression

We begin with a single CPWC acquisition that provides, for each steering angle $k \in \{1, \dots, K\}$, a beamformed in-phase and quadrature (I/Q) frame $F_k \in \mathbb{R}^{2\times H\times W}$. From the same acquisition, we also form a companion stack $\tilde{F}_k \in \mathbb{R}^{2\times H\times W}$ by high-pass filtering the same channel data (fourth-order Butterworth, cutoff 1.125 times the transmit center frequency) before beamforming, so that the companion frames emphasize the upper-band and harmonic content of the echoes. Because multipath reverberation is dominated by the fundamental band, whereas tissue harmonics build up along the direct propagation path [36], this harmonic-weighted companion stack serves as a physically motivated, reverberation-attenuated reference with finer axial resolution. The network is driven by $F_k$, whereas $\tilde{F}_k$ provides within-acquisition references, residual-derived negatives, and coherence-based self-supervisory cues [28].

During training, paired patches are sampled from both stacks with the same flip augmentation. Each stack is standardized independently per training instance to reduce gain and dynamic-range variation. For $X \in \{F, \tilde{F}\}$,

$$\mu_X = \mathbb{E}_{k,c,x,y}\left[X_{k,c}(x,y)\right], \tag{1}$$

$$\sigma_X = \sqrt{\mathbb{E}_{k,c,x,y}\left[\left(X_{k,c}(x,y) - \mu_X\right)^2\right]}. \tag{2}$$

$$\bar{X}_k = \frac{X_k - \mu_X}{\sigma_X + \epsilon_s}. \tag{3}$$

A reference angle $r$ is selected uniformly at random, and an encoder-decoder network $f_\theta$ produces the anchor prediction

$$S = f_\theta(\bar{F}_r) \in \mathbb{R}^{2\times H\times W}. \tag{4}$$

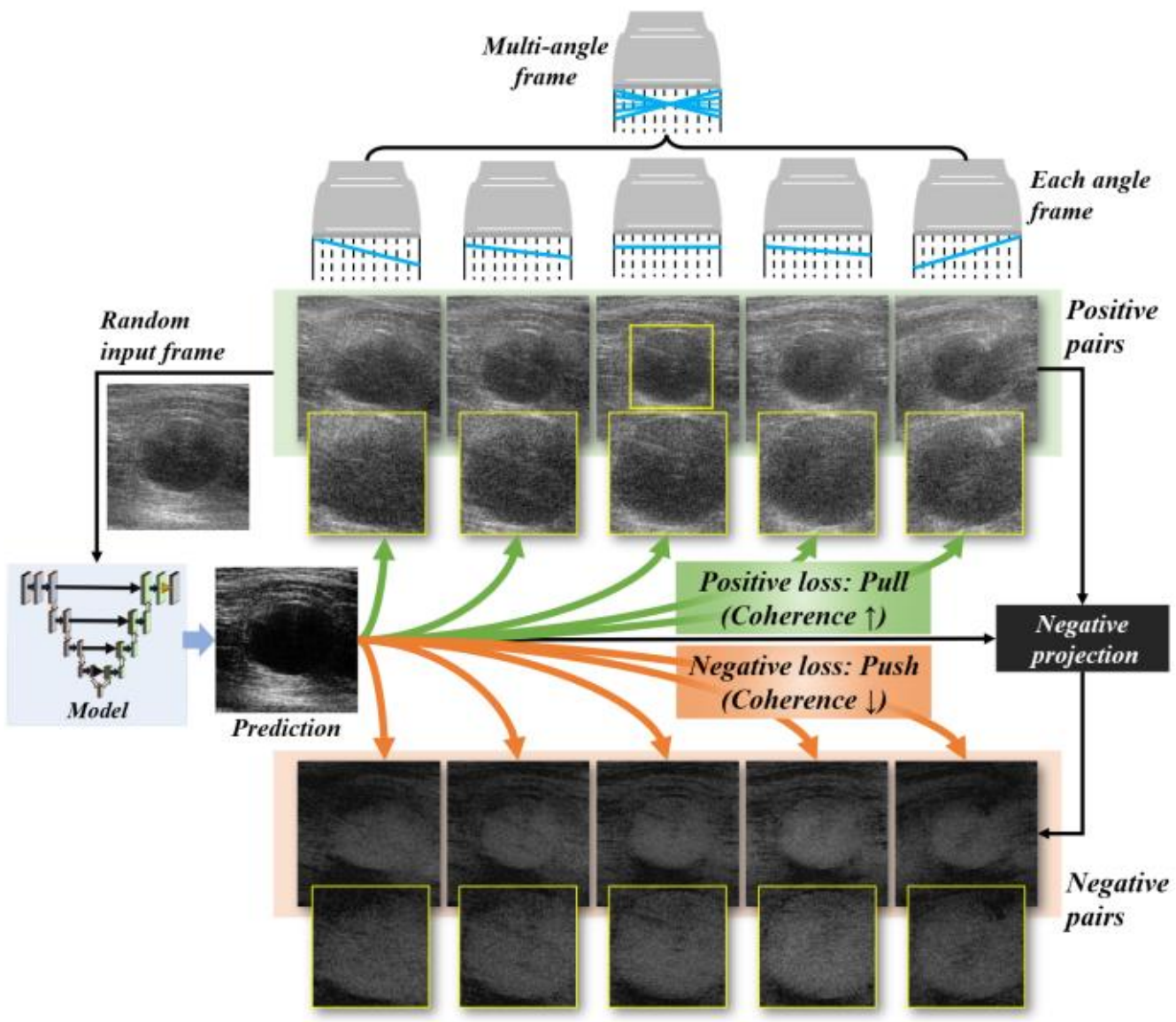

**Fig. 2** Overview of the proposed contrastive learning framework. A randomly selected single-angle frame is used as input, while the harmonic-weighted companion frames of all angles provide positive pairs. Residual projections from the prediction serve as negative pairs, and the network is trained to maximize coherence with positives and minimize coherence with negatives.

The random reference controls only the input view used to generate $S$; all $K$ companion angles are still used to define the self-supervisory keys. Consequently, the network never observes a compounded image during training and learns to suppress reverberation from a single steered transmission; at inference it can be applied either to a single-angle frame or to the coherent average of any number of angles (Section II-D).

The positive keys are the stop-gradient companion frames $P = \{\mathrm{sg}[\bar{\bar{F}}_k]\}_{k=1}^{K}$. To construct label-free negatives, we remove the anchor-aligned envelope component from each companion view while preserving its local phase structure. With $E(X) = \sqrt{X_I^2 + X_Q^2 + \epsilon_e}$,

$$E_k = E(\bar{\bar{F}}_k), \qquad E_S = E(S), \tag{5}$$

$$\alpha_k = \frac{\mathbb{E}_{x,y}[E_k E_S]}{\mathbb{E}_{x,y}[E_S^2] + \epsilon_p}, \qquad \Delta_k = E_k - \alpha_k E_S. \tag{6}$$

The residual mismatch is converted into a bounded weight and combined with the unit-magnitude phase field:

$$w_k = \exp(-|\Delta_k|/\beta), \tag{7}$$

$$\Phi_k = \frac{\bar{\bar{F}}_k}{E_k + \epsilon_e}, \qquad D_k = \mathrm{sg}[w_k \odot \Phi_k]. \tag{8}$$

Here, $D_k$ emphasizes steering-dependent mismatch, and the stop-gradient treatment makes the negative branch act only through gradients on $S$.

Anchor-key agreement is measured by a scale-aware cosine similarity. For $A, B \in \mathbb{R}^{2\times H\times W}$,

$$\rho(A,B) = \frac{\langle A, B\rangle}{\| A \|_2 \| B \|_2 + \epsilon_c}, \tag{9}$$

$$\eta(A,B) = \exp\left[-\gamma \left|\log\frac{\| A \|_2 + \epsilon_c}{\| B \|_2 + \epsilon_c}\right|\right], \tag{10}$$

$$s(A,B) = \rho(A,B)\eta(A,B). \tag{11}$$

The positive and negative similarities are $s_k^+ = s(S, \mathrm{sg}[\bar{\bar{F}}_k])$ and $s_k^- = s(S, D_k)$, yielding

$$L_{\mathrm{pos}} = \frac{1}{K}\sum_{k=1}^{K}(1 - s_k^+), \qquad L_{\mathrm{neg}} = \frac{1}{K}\sum_{k=1}^{K}|s_k^-|. \tag{12}$$

Thus, the anchor is pulled toward multi-angle references from the same acquisition while being decorrelated from residual-derived keys that capture angle-dependent clutter-like components.

We further use inter-angle coherence to distinguish phase-consistent anatomy from steering-dependent interference. From the companion complex field,

$$Z_k(x,y) = \bar{\bar{F}}_{k,I}(x,y) + j\bar{\bar{F}}_{k,Q}(x,y), \tag{13}$$

$$C(x,y) = \frac{|\sum_{k=1}^{K} Z_k(x,y)|}{\sum_{k=1}^{K}|Z_k(x,y)| + \epsilon_{\mathrm{coh}}}. \tag{14}$$

Low $C$ indicates angle-dependent responses, whereas high $C$ indicates stable phase across steering angles. Let

$$M_S = \sqrt{S_I^2 + S_Q^2 + \epsilon_m}, \qquad m_{\mathrm{zero}} = \mathbf{1}[C < \tau_{\mathrm{zero}}] \tag{15}$$

and $m_{\mathrm{keep}} = \mathbf{1}[C \geq \tau_{\mathrm{keep}}]$. The low-coherence weight and envelope floor are

$$\omega_{\mathrm{zero}} = \left(\delta + (1-\delta)(1-C)\right)^p m_{\mathrm{zero}}, \quad m_{\mathrm{floor}} = \min_{k,x,y} E_k. \tag{16}$$

The coherence-gated suppression and preservation losses are

$$L_{\mathrm{zero}} = \mathbb{E}_{\mathrm{batch}}\left[\frac{\sum |M_S - m_{\mathrm{floor}}| \omega_{\mathrm{zero}}}{\sum m_{\mathrm{zero}} + \epsilon}\right], \tag{17}$$

$$L_{\mathrm{keep}} = \mathbb{E}_{\mathrm{batch}}\left[\frac{\sum E_{\mathrm{ori}} m_{\mathrm{keep}}}{\sum m_{\mathrm{keep}} + \epsilon}\right], \tag{18}$$

$$E_{\mathrm{ori}} = \sqrt{\left(S_I - \bar{F}_{r,I}\right)^2 + \left(S_Q - \bar{F}_{r,Q}\right)^2 + \epsilon_m}. \tag{19}$$

$L_{\mathrm{zero}}$ softly lowers incoherent regions without enforcing exact zeros, whereas $L_{\mathrm{keep}}$ prevents over-suppression of coherent anatomical content.

The final objective is

$$L = L_{\mathrm{pos}} + L_{\mathrm{neg}} + \lambda_{\mathrm{zero}} L_{\mathrm{zero}} + \lambda_{\mathrm{keep}} L_{\mathrm{keep}}. \tag{20}$$

We use $\tau_{\mathrm{zero}} = 0.3$, $\tau_{\mathrm{keep}} = 0.7$, $\lambda_{\mathrm{zero}} = 0.025$, and $\lambda_{\mathrm{keep}} = 0.05$. The remaining constants are $\beta = 0.1$ (in standardized I/Q units), $\gamma = 1$, $\delta = 0.1$, and $p = 1.5$, and all $\varepsilon$ terms are $10^{-8}$ or smaller. For the positive-only ablation, $L_{\mathrm{neg}}$ and $L_{\mathrm{zero}}$ are removed. The model is optimized using AdamW with learning rate $10^{-4}$, weight decay $10^{-2}$, and an exponential moving average of the parameters with decay 0.999. At evaluation time, the input is either a single-angle frame or the coherent average of the available angles (all 75 angles unless stated otherwise); the input is standardized in the same manner, padded to a fixed multiple of 512, and cropped back to the original size.

Visualization and contrast metrics are computed from log-compressed magnitude images with a fixed dynamic range.

### *C. Network architecture and training strategies*

Fig. 3 illustrates the designed backbone which is a symmetric encoder-decoder network with skip concatenations that maps a single two-channel I/Q frame to a two-channel prediction on the same beamforming grid, $f_\theta: \mathbb{R}^{2\times H\times W} \rightarrow \mathbb{R}^{2\times H\times W}$. The U-Net-style backbone, termed SonoNeXt, combines ConvNeXt-style local spatial modeling with transformer-style global context aggregation.

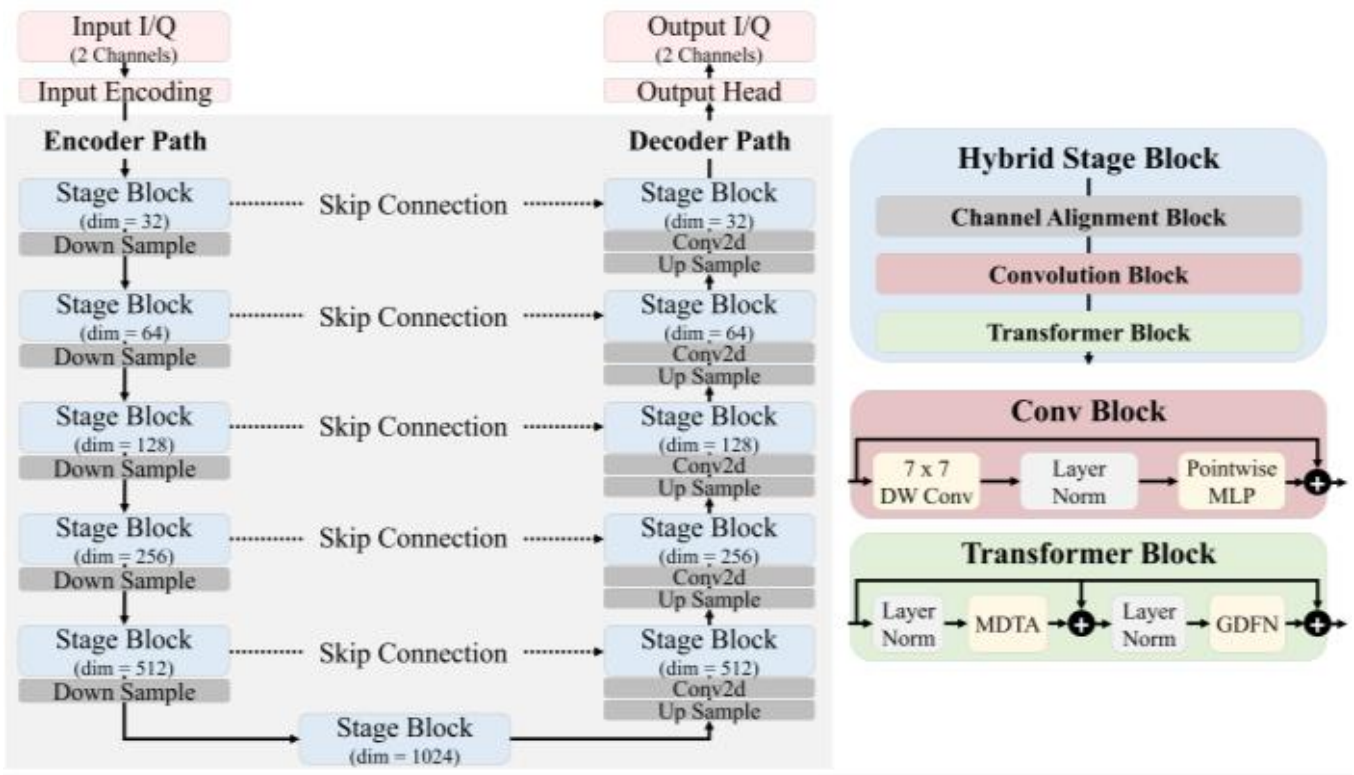

**Fig. 3** SonoNeXt encoder-decoder backbone. The network maps a two-channel I/Q input to a two-channel I/Q prediction using a six-stage U-Net architecture with skip connections, ConvNeXt-style local blocks, and transformer-style global blocks.

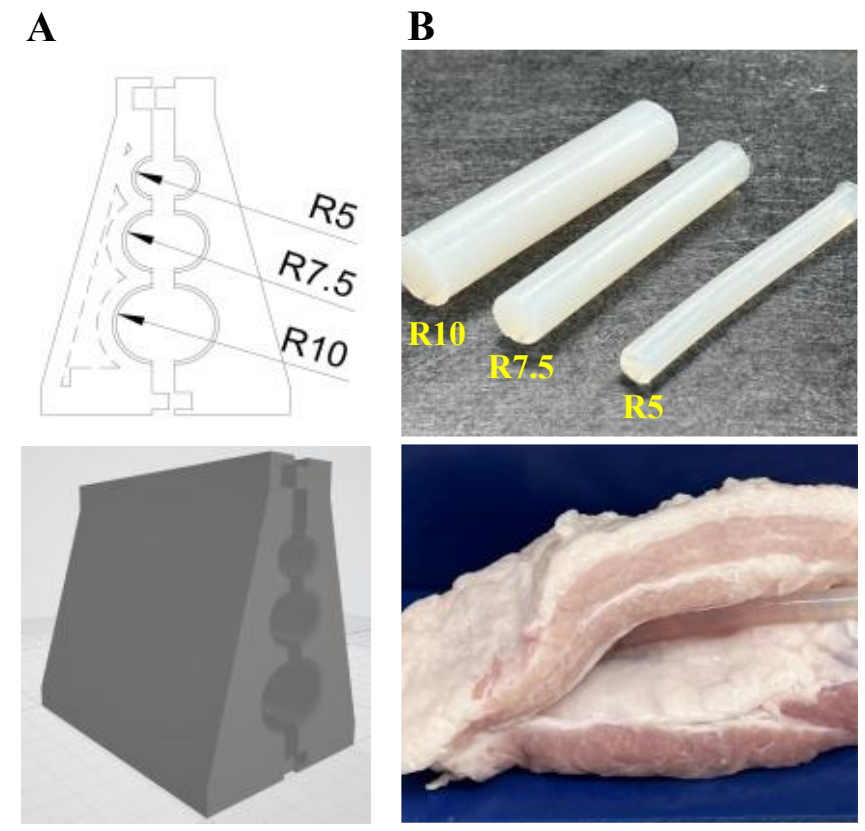

**Fig. 4** (A) Schematic and 3D-printed phantom mold with agar–cellulose anechoic targets of radii 5, 7.5, and 10 mm. (B) Matching silicone rods and *ex vivo* pork belly were used to generate realistic reverberation.

Given an input I/Q frame $x \in \mathbb{R}^{2\times H\times W}$, the network first lifts the two channels to the Stage-1 width (32) using a $1\times 1$ channel-alignment layer (Conv-GroupNorm-LeakyReLU) embedded in the first Hybrid Stage Block. The encoder then proceeds through $L = 6$ stages with channel widths ($d_1$, …, $d_6$) = (32, 64, 128, 256, 512, 1024). Between the consecutive encoder stages, a stride- $2\times 2$ convolution halves the spatial resolution and increases the channel width. Each stage applies a Hybrid Stage Block that combines local spatial modeling and global context aggregation: a ConvNeXt-style depthwise $7\times 7$ convolution with GroupNorm and a pointwise MLP (expand-LeakyReLU-project), followed by a transformer-style block consisting of MDTA (multi-deconv head transposed attention) and GDFN (gated depthwise feed-forward network), both wrapped in pre-normalization (GroupNorm) and residual connections.

The decoder mirrors the encoder. At each decoder level, features are upsampled by bilinear interpolation and projected to the target stage width using a $1\times 1$ convolution, then concatenated with the corresponding encoder features through skip connections. The concatenated tensor is fused at the beginning of the decoder stage by a $1\times 1$ channel-alignment layer and further refined by the same Hybrid Stage Block. Finally, a $3\times 3$ convolutional head maps the highest-resolution features back to two channels to produce the predicted I/Q output $S \in \mathbb{R}^{2\times H\times W}$.

### D. Experimental setups

#### *1) Dataset and acquisition protocol*

The dataset consists of a custom phantom and public *in vivo* scans. The phantom is a tissue-mimicking agar–cellulose block cast in a 3D-printed mold with three cylindrical cavities of radius 5, 7.5, and 10 mm (Fig. 4A), which serve as anechoic targets for quantitative assessment. Silicone rods of matching radii were inserted as strong specular reflectors that promote multi-bounce propagation (Fig. 4B), and a slab of *ex vivo* pork belly placed between the probe and the phantom adds layered interfaces that generate strong, clinically representative reverberation overlapping the targets (DAS column of Fig. 5).

For all acquisitions, raw RF channel data were recorded using a research ultrasound system (Vantage 128, Verasonics Inc., Redmond, WA, USA) equipped with an L11-5v linear-array transducer. Each sequence employed CPWC with $K = 75$ steering angles uniformly spanning $-18°$ to $+18°$ (including broadside) at a fixed angular step. To avoid axial blurring, no baseband demodulation was applied: for each steering angle, the channel data were beamformed with delay-and-sum (DAS) and receive apodization, and the analytic signal of the beamformed radio-frequency data was retained as the I/Q frame. A Cartesian beamforming grid with pixel spacing of $\lambda_f/5$ was used, where $\lambda_f = c/f_c$. For training, each acquisition was standardized to zero mean and unit variance over I/Q, random $512 \times 512$ crops were extracted, and random horizontal flips were applied. In total, 356 CPWC acquisitions were collected for the custom reverberation dataset; 325 acquisitions (24,375 single-angle frames) were used for training and 31 acquisitions were held out for validation. The public PICMUS [37] and CUBDL [38-40] *in vivo* datasets were used only for evaluation, as summarized in Table I.

#### *2) Baseline methods*

Baselines were DAS, DMAS, the supervised 3D CNN of [12] with the authors' pretrained weights (its full-synthetic-aperture input was obtained by retrospectively encoding each PW stack into virtual transmit events from the element geometry and steering delays [39, 40]), and DCL, an unsupervised beamformer trained to maximize inter-angle coherence, also with the authors' pretrained weights [28]. pCoRev was trained on the same training split without labels (Sections II-B and II-C). For the main comparison (Sections III-A to III-C), all methods received the same 75-angle CPWC acquisitions, and the pCoRev input was the coherent average of all angles, i.e.,

TABLE I
SUMMARY OF THE TRAINING AND VALIDATION DATASETS. EVERY ACQUISITION IS A 75-ANGLE CPWC SET (K = 75); THE 325 TRAINING SETS CORRESPOND TO 24,375 SINGLE-ANGLE FRAMES.

| Dataset | Category | Sets (train / val) | Ref. |
|---|---|---|---|
| REVERB (ours) | *Ex vivo* phantom | 325 / 31 | This work |
| PICMUS | *In vivo* carotid | 0 / 1 | [37] |
| CUBDL (JHU) | *In vivo* breast | 0 / 1 | [38-40] |

TABLE II

QUANTITATIVE COMPARISON ON THE 31 PHANTOM VALIDATION ACQUISITIONS (75-ANGLE CPWC INPUT; MEAN ± SD). ΔB: MEAN BACKGROUND LEVEL RELATIVE TO DAS; SSNR: BACKGROUND SPECKLE SNR; SSIM: SIMILARITY TO DAS OUTSIDE THE TARGET. BEST CR, CNR, AND GCNR IN BOLD.

| Metric | DAS | DMAS | 3D CNN | DCL | pCoRev |
|---|---|---|---|---|---|
| CR (dB) | 20.45 ± 3.84 | 24.37 ± 6.02 | 30.89 ± 7.65 | 27.47 ± 5.73 | **38.76 ± 5.40** |
| CNR (dB) | 8.38 ± 2.00 | 6.92 ± 2.79 | 9.38 ± 3.28 | 9.91 ± 2.37 | **11.75 ± 1.81** |
| gCNR | 0.832 ± 0.085 | 0.754 ± 0.142 | 0.826 ± 0.103 | 0.878 ± 0.089 | **0.932 ± 0.051** |
| sSNR | 1.12 ± 0.20 | 0.72 ± 0.15 | 0.98 ± 0.20 | 1.05 ± 0.22 | 0.80 ± 0.16 |
| ΔB (dB) | 0.0 ± 0.0 | -14.5 ± 4.0 | -4.2 ± 2.5 | -2.0 ± 2.2 | -8.2 ± 3.1 |
| SSIM | 1.00 ± 0.00 | 0.28 ± 0.04 | 0.24 ± 0.05 | 0.49 ± 0.05 | 0.32 ± 0.03 |

the DAS image itself, so that any improvement is attributable to the network. To assess ultrafast operation (Section III-D), the number of angles available at test time was reduced to $K_{\text{test}} \in \{1, 3, 5, 7, 9, 11, 15, 19, 25, 37, 75\}$: the $K_{\text{test}}$ angles were selected uniformly over the full ±18° span (broadside only for $K_{\text{test}} = 1$), coherently averaged, and fed to the same trained network without retraining, and DAS with the same $K_{\text{test}}$ angles served as the reference.

### 3) Evaluation metrics

Quantitative image quality was assessed using the contrast ratio (CR), contrast-to-noise ratio (CNR), and generalized contrast-to-noise ratio (gCNR), measured with identical masks across methods. Following common practice, regions were defined as an isoechoic background ($i$) and an anechoic target ($o$). CNR (in dB) was computed as

$$\text{CNR(dB)} = 20\log_{10}\left(\frac{|\mu_i - \mu_o|}{\sqrt{\frac{(\sigma_i^2 + \sigma_o^2)}{2}}}\right), \qquad (21)$$

where $\mu$ and $\sigma$ denote the mean and standard deviation within each region. The gCNR was computed as

$$\text{gCNR} = 1 - \int \min\{d_i(x), d_o(x)\}dx \qquad (22)$$

where $d_i$ and $d_o$ are the probability density functions of pixel intensities in the isoechoic and anechoic regions, respectively.

To quantify the residual reverberation level inside the anechoic target directly, the CR was computed on the envelope (linear-amplitude) image as

$$\text{CR(dB)} = 20\log_{10}\left(\frac{\hat{\mu}_i}{\hat{\mu}_o}\right), \qquad (23)$$

where $\hat{\mu}_i$ and $\hat{\mu}_o$ denote the mean envelope amplitudes of the isoechoic background and the anechoic target, respectively. Unlike CNR and gCNR, CR is computed on linear amplitudes and is insensitive to the within-region variance, so it isolates the depth of suppression. As an auxiliary check of texture preservation, we also report the speckle signal-to-noise ratio of the background, $\text{sSNR} = \frac{\hat{\mu}_i}{\hat{\sigma}_i}$, where $\hat{\sigma}_i$ is the standard deviation of the background envelope [41]; the change of the mean background level relative to DAS after peak normalization, ΔB, which indicates whether diffuse tissue signal is retained; and the structural similarity index (SSIM) [42] between each output and the DAS image computed outside the anechoic target. All metrics were computed with identical regions of interest for every method and are summarized in Table II.

## III. RESULTS

### A. Phantom study

Figs. 5 and 6 show representative validation cases with 75-angle CPWC input for every method (used throughout Sections III-A to III-C): cross-sectional targets of radius 5, 7.5, and 10 mm in Fig. 5 and the corresponding longitudinal targets in Fig. 6, each with magnified views of the target region. In the DAS images, reverberation fills the anechoic targets and blurs their boundaries. DMAS raises the overall contrast but leaves residual clutter and nonuniform darkening, particularly below the target. The supervised 3D CNN suppresses clutter strongly but produces patchy texture and locally distorted boundaries, and DCL preserves the target shape but leaves residual haze. pCoRev yields the darkest and most homogeneous target interior with a clear boundary and reduced posterior haze in all four cases, including the longitudinal targets, where the artifact overlaps the target over a broad axial extent.

The axial profiles (last column of Figs. 5 and 6) confirm these observations: pCoRev produces the deepest and most stable trough across the target span while recovering the intensity outside the target as sharply as DAS, whereas the 3D CNN profile shows abrupt drops indicative of over-suppression, and DMAS and DCL retain a higher residual level inside the target.

Table II summarizes the quantitative performance on the 31 validation acquisitions with all 75 angles. pCoRev achieved the highest CR, CNR, and gCNR, with a CR of 38.76 ± 5.40 dB, a CNR of 11.75 ± 1.81 dB, and a gCNR of 0.932 ± 0.051. In comparison, the strongest competing baselines reached 30.89 ± 7.65 dB in CR (3D CNN), 9.91 ± 2.37 dB in CNR (DCL), and 0.878 ± 0.089 in gCNR (DCL). All improvements were significant in paired Wilcoxon signed-rank tests against every baseline ($p < 10^{-4}$, $n = 31$): pCoRev exceeded DAS, DMAS, and DCL in CR and CNR on all 31 acquisitions and exceeded the 3D CNN on 28 (CR) and 26 (CNR) of them.

The 18 dB gain in CR over DAS corresponds to a residual

TABLE III

*IN VIVO* RESULTS OVER FOUR CLINICAL ACQUISITIONS (PICMUS CAROTID, LONGITUDINAL AND CROSS-SECTIONAL [37]; CUBDL JHU BREAST, ACQUISITIONS 28 AND 32 [38]; 75-ANGLE CPWC INPUT): MEAN ± SD ACROSS ACQUISITIONS AND MEAN RANK OVER THE 12 CASE–METRIC COMBINATIONS (1 = BEST). BEST VALUE PER ROW IN BOLD.

| Metric | DAS | DMAS | 3D CNN | DCL | pCoRev |
|---|---|---|---|---|---|
| CR (dB) | 19.4 ± 2.6 | 24.0 ± 4.5 | 32.7 ± 3.5 | 25.4 ± 7.7 | **34.6 ± 6.2** |
| CNR (dB) | 9.5 ± 1.7 | 9.1 ± 1.2 | 10.1 ± 1.7 | 10.5 ± 1.8 | **12.4 ± 2.4** |
| gCNR | 0.90 ± 0.05 | 0.87 ± 0.07 | 0.92 ± 0.04 | 0.91 ± 0.06 | **0.96 ± 0.03** |
| Mean rank | 4.1 | 3.7 | 2.5 | 3.1 | **1.7** |

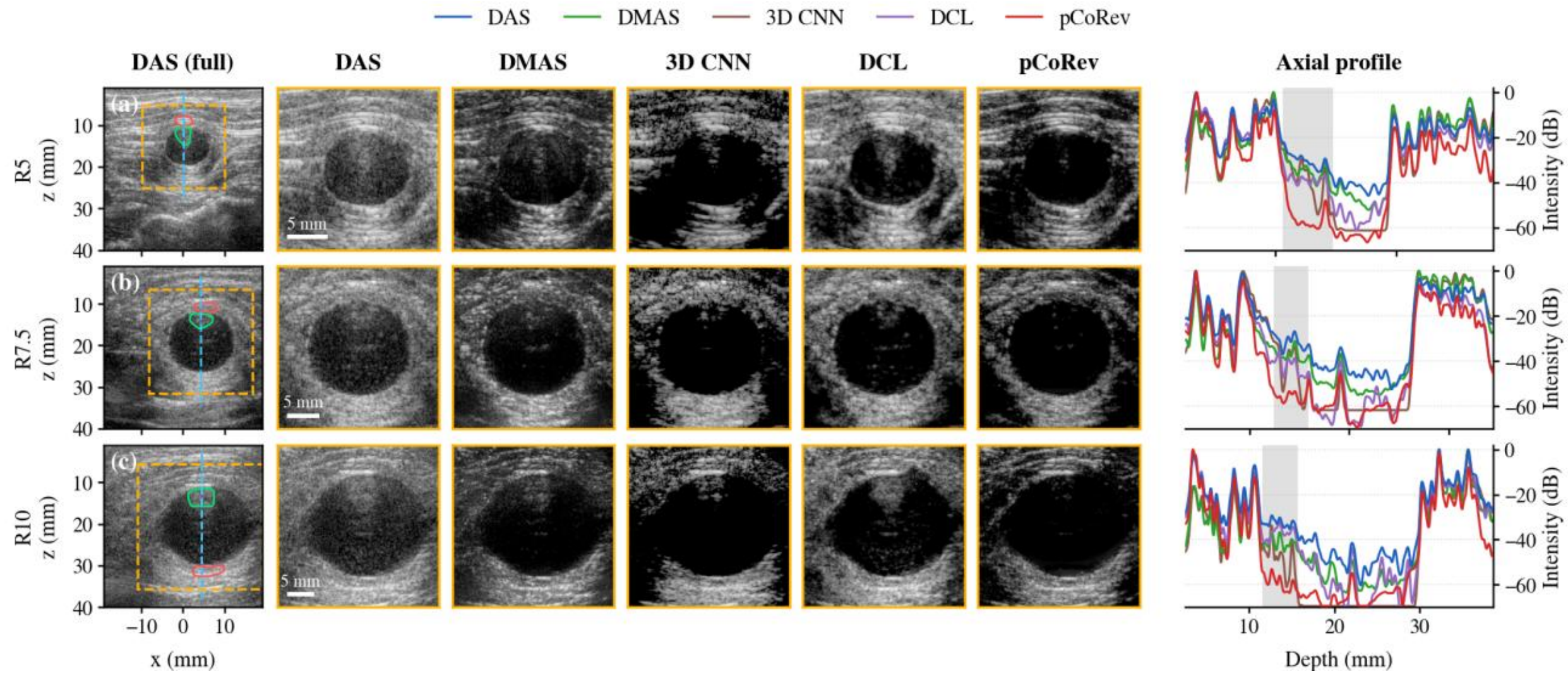

**Fig. 5** Cross-sectional phantom targets of radius (a) 5, (b) 7.5, and (c) 10 mm with 75-angle CPWC input for all methods. First column: DAS image with the target (green) and background (red) ROIs, the axial-profile line (dashed) and the magnified region (dashed box); columns 2–6: magnified views for each method (scale bar: 5 mm); last column: axial profiles along the dashed line (envelope averaged over a 1-mm-wide lateral strip; shaded band: depth range of the target ROI); 70 dB dynamic range.

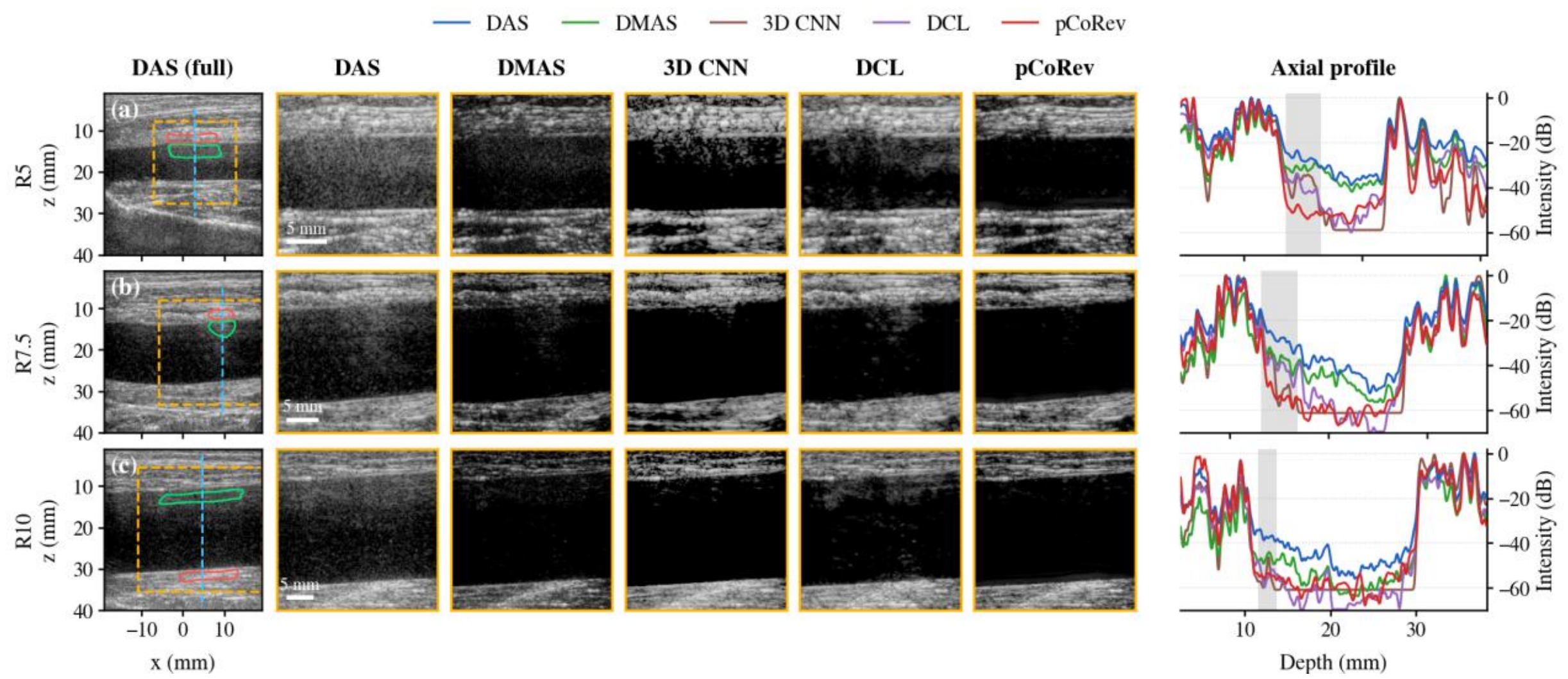

**Fig. 6** Longitudinal phantom targets of radius (a) 5, (b) 7.5, and (c) 10 mm with 75-angle CPWC input for all methods; layout as in Fig. 5.

clutter amplitude inside the target that is about eight times lower, and the simultaneous gains in CNR and gCNR indicate improved separability of the target and background distributions. The preservation-oriented metrics in Table II show the cost of this suppression: the background speckle SNR of pCoRev (0.80 ± 0.16) was lower than that of DAS (1.12 ± 0.20) and DCL (1.05 ± 0.22) but higher than that of DMAS (0.72 ± 0.15); its mean background level was 8.2 ± 3.1 dB below DAS (DMAS: −14.5 dB; 3D CNN: −4.2 dB; DCL: −2.0 dB); and its SSIM with respect to DAS outside the target (0.32 ± 0.03) lay between those of DCL (0.49 ± 0.05) and the 3D CNN (0.24 ± 0.05). pCoRev thus alters speckle texture and attenuates diffuse tissue signal more than DCL but less than DMAS and the channel-domain 3D CNN (Section IV-B).

### *B. In vivo study*

Fig. 7 shows in vivo results for (a) a PICMUS carotid acquisition and (b) a CUBDL breast acquisition. DAS and DMAS retain residual clutter in the vessel lumen and the cyst, the 3D CNN suppresses it aggressively at the cost of texture, and DCL leaves residual haze, whereas pCoRev yields a darker, more homogeneous low-signal region with clear tissue interfaces and the deepest in-target profile. Table III summarizes CR, CNR, and gCNR over four in vivo acquisitions (PICMUS carotid, longitudinal and cross-sectional; CUBDL JHU breast, acquisitions 28 and 32), none of which were used for training. pCoRev obtained the highest mean value in all three metrics (CR 34.6 dB, CNR 12.4 dB, gCNR 0.96) and the best mean rank (1.7 versus 2.5 for the 3D CNN, the strongest baseline); it was the best or second-best method in 10 of the 12 case–metric combinations and never fell below DAS, whereas every baseline ranked fourth or fifth in at least one acquisition. The 3D CNN, the closest competitor in CR, reaches its values partly through signal collapse inside the target (the flat profile floors in Fig. 7), which CR rewards but gCNR and the preservation metrics in Table II penalize. pCoRev improved CR over DAS by 9–21 dB in every acquisition, indicating that the suppression learned on the phantom transfers to other scanners and probes without fine-tuning.

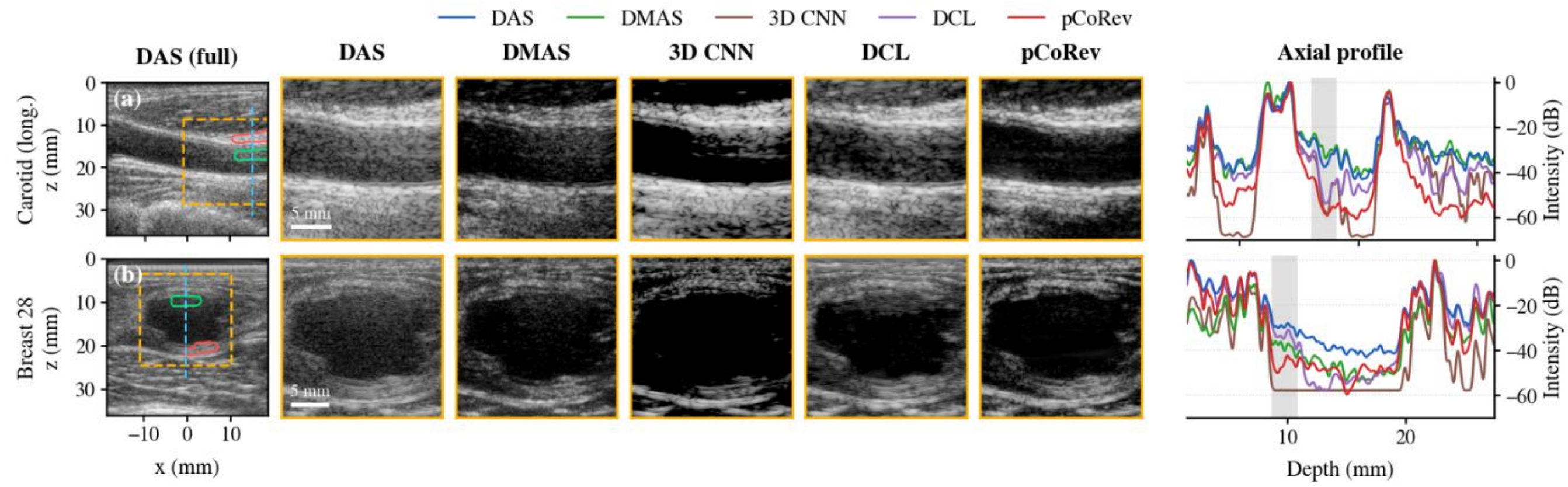

**Fig. 7** *In vivo* results with 75-angle CPWC input: (a) PICMUS carotid artery, longitudinal view; (b) CUBDL JHU breast acquisition 28; layout as in Fig. 5 (magnified region: vessel lumen and cyst).

## C. Ablation study

To assess the pull-push training strategy, we compared the full pCoRev objective with a positive-only variant that removes the residual-derived negative and low-coherence suppression terms. As shown in Fig. 8, the full model converged rapidly to a validation CNR of approximately 11.8, whereas the positive-only variant peaked early and settled near 9.0. This gap indicates that positive alignment alone is insufficient and that the residual-derived negative and coherence-gated suppression terms are essential for robust reverberation reduction.

## D. Reverberation suppression with reduced transmit angles

Fig. 9 examines how pCoRev behaves when the number of angles available at test time, $K_{\text{test}}$, is reduced from 75 to one, i.e., when the achievable frame rate is increased by up to 75-fold. For DAS, CR, CNR, and gCNR degraded steadily as fewer angles were compounded (Fig. 9(a) and (b)), from 20.45 ± 3.84 dB, 8.38 ± 2.00 dB, and 0.832 ± 0.085 at $K_{\text{test}} = 75$ to 14.94 ± 5.12 dB, 5.75 ± 3.71 dB, and 0.711 ± 0.181 for a single broadside transmission. pCoRev was markedly more robust: with a single transmission it still achieved a CR of 32.10 ± 7.79 dB, a CNR of 10.39 ± 3.02 dB, and a gCNR of 0.872 ± 0.106, exceeding 75-angle DAS in all three metrics (paired Wilcoxon tests, $p < 0.01$, $n = 31$; higher CR in 31 and higher CNR in 28 of the 31 acquisitions) and 75-angle DCL in CR ($p < 10^{-4}$), whereas its CNR and gCNR were statistically indistinguishable from those of 75-angle DCL (p = 0.19 and 0.51). With $K_{\text{test}} = 7$, pCoRev reached 98% of its 75-angle CNR (11.50 ± 1.74 dB). The dip at $K_{\text{test}} = 3$, visible for both methods, arises because this subset consists of the broadside and the two extreme (±18°) angles, whose transmissions have the lowest signal-to-noise ratio. The *in vivo* cases followed the same trend: with a single transmission, pCoRev retained a CNR of 11.10 dB (carotid) and 12.68 dB (breast), compared with 8.74 and 7.99 dB for single-angle DAS and 11.51 and 8.64 dB for 75-angle DAS in the same pipeline. Fig. 9(c)–(f) illustrate this on a cross-sectional R5 phantom: the target is barely discernible in the single-angle DAS image, whereas single-angle pCoRev recovers a well-delineated target comparable to the 75-angle result. pCoRev can therefore deliver reverberation-suppressed images at the frame rate of single-transmission PW imaging.

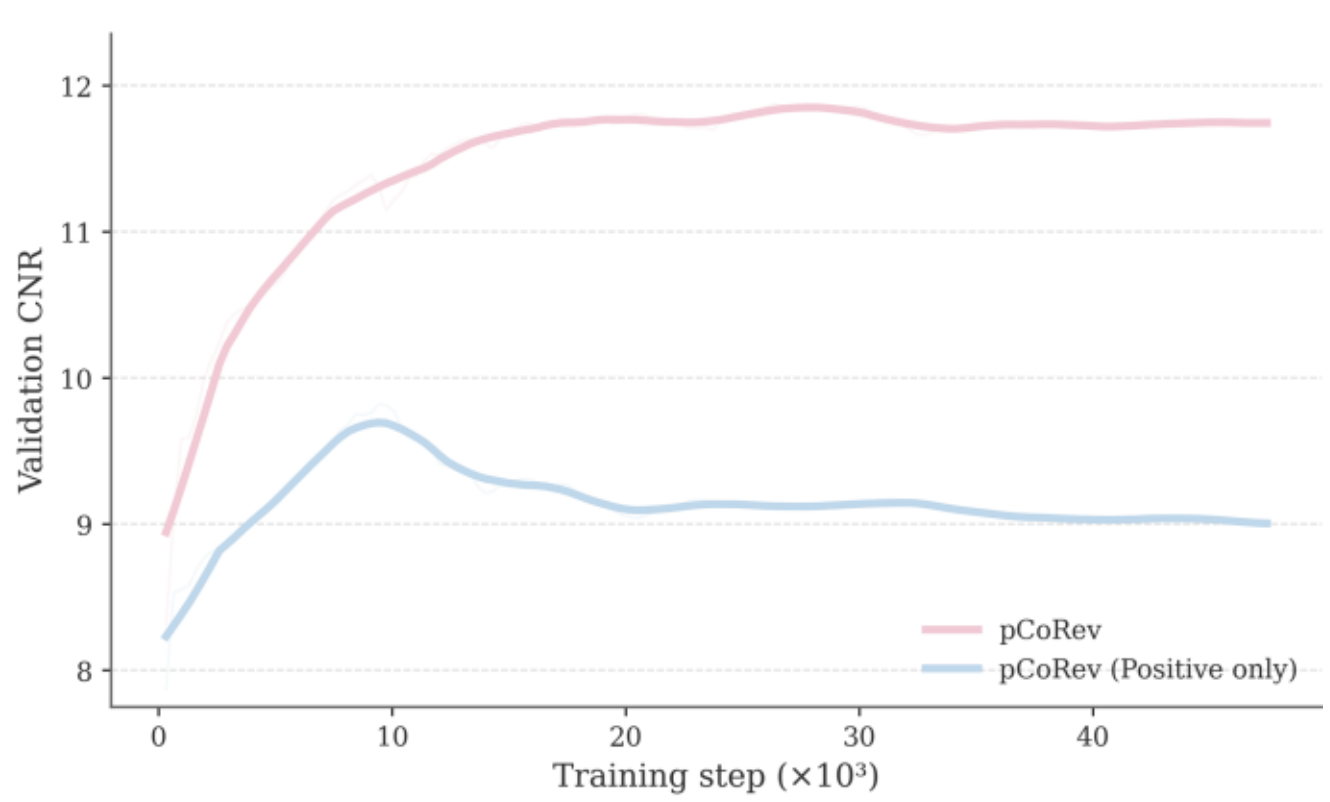

**Fig. 8** Ablation study on the reverberation validation dataset. The full pCoRev objective converges to a higher and more stable validation CNR than the positive-only variant.

# IV. Discussion

## A. Benefits of our framework

pCoRev converts the angle dependence of multipath reverberation into a self-supervised learning signal: within-acquisition angular consistency provides positive references, residual projections provide negative cues, and coherence gating regularizes suppression and preservation, so no artifact-free targets are needed. The phantom, *in vivo*, and ablation results indicate that the improvement stems from this reverberation-specific pull-push design rather than from generic self-supervised learning alone.

A distinctive property of pCoRev is that the network is trained on, and applied to, a single steered transmission; compounding is used only to construct the self-supervisory keys. This decouples reverberation suppression from the number of transmissions: a single transmission processed by pCoRev yields a higher CR and CNR than 75-angle CPWC (Section III-D), i.e., a 75-fold reduction in transmit events and a frame rate bounded by the round-trip time of one plane wave rather than by the compounding sequence. The method is therefore directly applicable to ultrafast modalities such as shear-wave elastography and ultrafast Doppler, in which compounding is limited by the required temporal resolution and reverberation from superficial layers is a major source of clutter [17]. Two further properties support its generality: the harmonic-weighted companion stack adds a second physical prior, because multipath echoes are dominated by the fundamental band and the companion keys are therefore intrinsically less contaminated by reverberation; and pCoRev operates on beamformed I/Q data, so it can follow any beamformer without channel-data access, unlike channel-domain approaches such as the 3D CNN [12]. Consistently, the *in vivo* acquisitions came from different systems and probes than the training phantom data [37], [38] and were processed without fine-tuning.

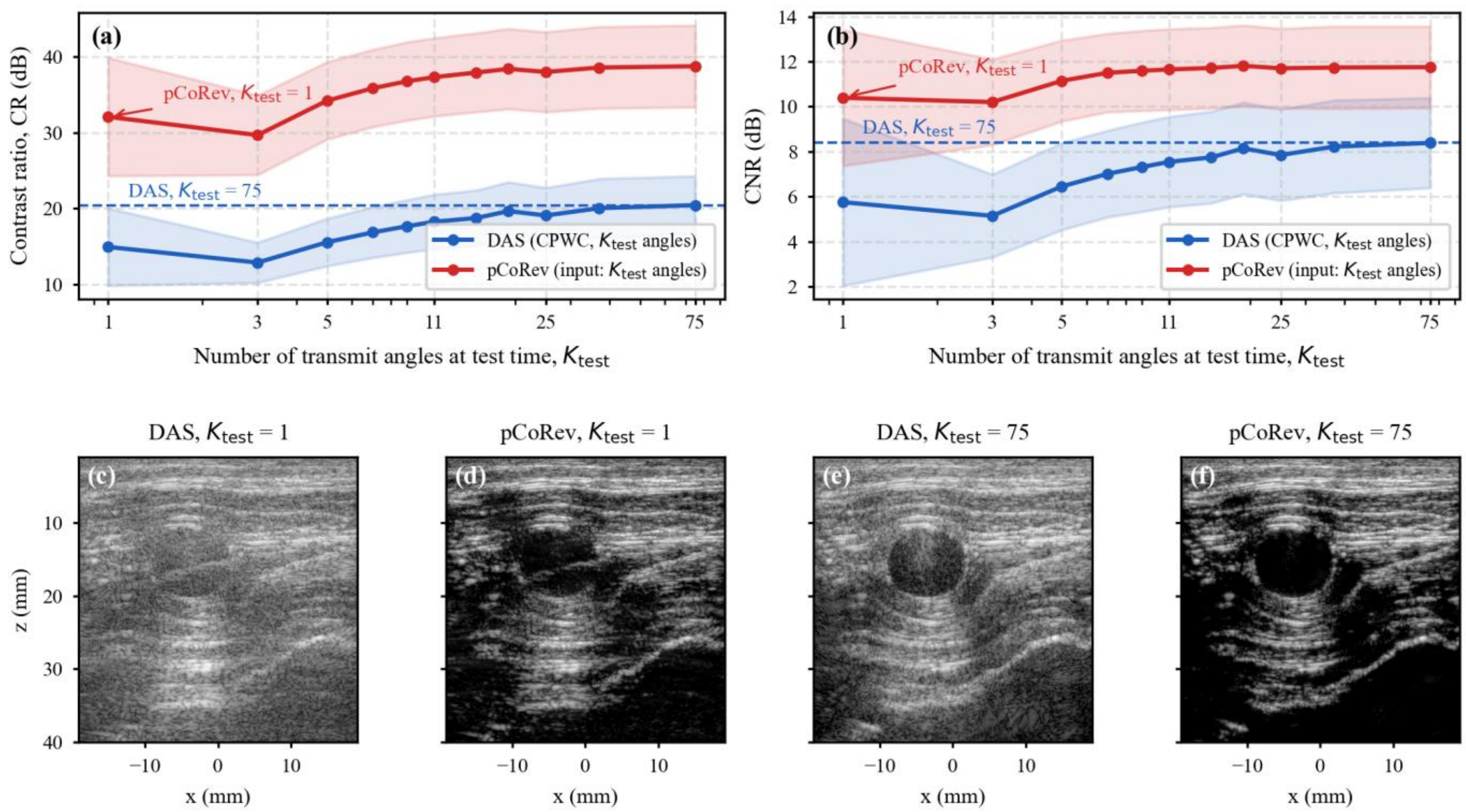

**Fig. 9** Reverberation suppression versus the number of transmit angles $K_{test}$ used at test time (31 phantom validation acquisitions; solid lines: mean, shaded: ± one standard deviation). (a) CR and (b) CNR of DAS compounded over $K_{test}$ angles and of pCoRev fed with the same $K_{test}$-angle average; the dashed line marks 75-angle DAS. (c)–(f) A cross-sectional R5 phantom case reconstructed from a single broadside transmission (c, d) and from all 75 angles (e, f) with DAS (c, e) and pCoRev (d, f); 70 dB dynamic range.

### *B. Limitations and future studies*

Several limitations remain. First, the custom phantom with *ex vivo* tissue is more realistic than simulation but still a controlled setting, and the *in vivo* evaluation is limited in scale; broader validation across organs, scanning conditions, and patients is required. Second, not all auxiliary metrics favored pCoRev: the background speckle SNR decreased relative to DAS (0.80 versus 1.12) and the mean background level was about 8 dB lower, indicating that stronger suppression is accompanied by a change in speckle statistics and a mild attenuation of diffuse tissue signal; extending the coherence-gated preservation term with a speckle-statistics prior is a natural remedy. Third, the 3D CNN and DCL baselines were used with the weights released by their authors; retraining them on the same phantom acquisitions could narrow the reported gap.

Fourth, the implementation is not yet optimized for real-time use: one 1536 × 1536 I/Q frame takes about 0.78 s on an NVIDIA RTX 4090 GPU (0.32 s for 1024 × 1024) in single-precision PyTorch with 36 M parameters. Lightweight architectures, pruning, quantization, and hardware-aware implementation are needed, although single-transmission operation already removes the acquisition-time penalty of compounding.

Finally, the present evaluation mainly relies on image-quality metrics and visual inspection. Since CR, CNR, and gCNR do not fully capture clinical utility, future studies should include task-oriented evaluations such as lesion conspicuity assessment, reader studies, and downstream diagnostic or measurement tasks.

## V. Conclusion

This study presented pCoRev, an unsupervised physics-informed contrastive learning framework that suppresses reverberation artifacts in ultrafast PW imaging by exploiting the angular coherence differences between tissue echoes and multipath reverberation, without paired artifact-free references. The pull-push objective with coherence-gated regularization achieved the best contrast metrics on phantom and *in vivo* data, and, because the network operates on single steered transmissions, a single plane wave processed by pCoRev outperformed 75-angle compounding. Future work will address broader clinical validation and real-time optimization.